\documentclass[sigconf]{acmart}
\AtBeginDocument{%
  }

\copyrightyear{2025} 
\acmYear{2025} 
\setcopyright{cc}
\setcctype{by-nc-nd}
\acmConference[MEMOCODE '25]{International Symposium on Formal Methods and Models for System Design}{September 28-October 3, 2025}{Taipei, Taiwan}
\acmBooktitle{International Symposium on Formal Methods and Models for System Design (MEMOCODE '25), September 28-October 3, 2025, Taipei, Taiwan}\acmDOI{10.1145/3742875.3754678}
\acmISBN{979-8-4007-1994-3/2025/09}

\usepackage{multirow}

\usepackage{subcaption} 
\begin{document}

\title{Coherence-Aware Task Graph Modeling for Realistic Application}



\author{Guochu Xiong}

\affiliation{%
  \institution{College of Computing and Data Science, Nanyang Technological University}
  \country{Singapore}
}

\author{Xiangzhong Luo}
\affiliation{%
  \institution{School of Computer Science and Engineering, Southeast University}
   \country{China}
}

\author{Weichen Liu}
\authornote{
Corresponding author (email: liu@ntu.edu.sg).\\
This work is partially supported by the Ministry of Education, Singapore, under its Academic Research Fund Tier 1 (RG94/23) and Tier 2 (MOE-T2EP20224-0006).}
\affiliation{%
  \institution{College of Computing and Data Science, Nanyang Technological University}
  \country{Singapore}
}


\begin{abstract}

 As multicore systems continue to scale, cache coherence has emerged as a critical determinant of system performance, with coherence behavior and task execution deeply intertwined—reshaping inter-task dependencies. Task graph modeling offers a structured way to capture such dependencies and serves as the foundation for many system-level design strategies. However, these strategies typically rely on predefined task graphs, while many real-world applications lack explicit task graphs and exhibit dynamic, data-dependent behavior, limiting the effectiveness of static approaches. Thus, many task graph modeling methods for realistic workloads have been developed. However, they either rely on implicit methods—using application-specific features without producing explicit graphs—or generate graphs tailored to fixed scheduling models, offering limited generality. Critically, they overlook coherence interactions, resulting in a mismatch between design assumptions and actual runtime behavior. To address these limitations, we propose CoTAM, a Coherence-Aware Task Graph Modeling framework for realistic workloads that constructs a unified task graph reflecting runtime behavior. CoTAM analyzes the impact of coherence by decoupling its behavior from overall execution. It then quantifies the influence of coherence through a learned weighting scheme, infers inter-task dependencies for coherence-aware task graph generation. Extensive experiments demonstrate the superiority of CoTAM over implicit methods, not only bridging the gap between dynamic workload behavior and existing designs, but also underscoring the importance of incorporating cache coherence into task graph modeling for accurate and generalizable system-level analysis.

 
\end{abstract}






\maketitle

\section{Introduction}
\label{introduction}
As modern computing systems grow in complexity, cache coherence has become a critical factor influencing system performance. To ensure data consistency across cores, coherence protocols such as MESI \cite{b10} are employed to manage operations such as cache-to-cache transfers and invalidations. For example, when one core modifies a shared cache line, coherence actions are triggered to update or invalidate copies in other cores. While essential for correctness, these operations introduce considerable overhead—including increased traffic, latency, and energy consumption. Due to its wide-reaching impact, cache coherence has been extensively studied in both application-level contexts and architectural domains such as Network-on-Chip (NoC) design, where coherence-induced communication significantly influences system behavior \cite{b22}.

\begin{figure}[htbp]
    \centering
        \includegraphics[width=1.01\linewidth]{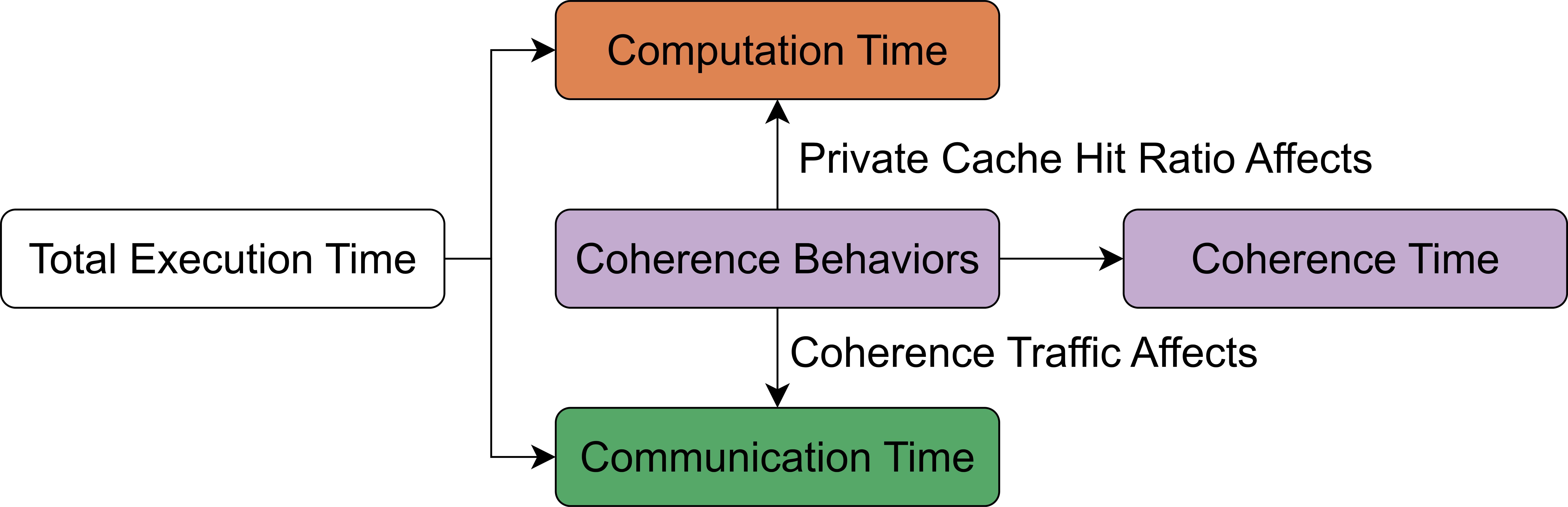}

    \caption{Impact of coherence behaviours on task execution}
    \label{fig:coherence_impact}
\end{figure}

Although coherence mechanisms operate at the architectural level, they are fundamentally triggered by how tasks access and share data—revealing implicit inter-task dependencies. To better understand and quantify the impact of coherence behavior, we decouple coherence overhead from conventional computation and communication costs. As illustrated in Figure~\ref{fig:coherence_impact}, task execution time comprises both computation and communication phases. Coherence behavior influences computation through the private cache hit ratio: a higher hit ratio implies more data served locally, reducing remote accesses and latency, while coherence misses lead to stalls and increased delay. On the communication side, coherence-related activities—such as cache-to-cache transfers and invalidation messages—contribute to network traffic, elevate contention, and further degrade performance. These delays are collectively represented as coherence time, a key metric reflecting the performance cost of maintaining memory consistency.

Thus, coherence behavior both originates from and influences task execution; coherence-induced delays directly affect when and how tasks execute, particularly when multiple tasks actively contend for shared data. These delays can introduce implicit synchronization and reshape critical inter-task dependencies. To effectively manage and optimize such effects, it is essential to explicitly model how coherence behavior translates into inter-task dependencies during execution.

Task graph modeling provides a structured way to capture such dependencies by representing communication and coordination patterns among tasks. It has become foundational in many system-level design strategies, guiding critical decisions such as task mapping and scheduling. This is especially evident in existing work \cite{b2,b3,b9, b23} that relies on predefined, static task graphs (e.g., \cite{b21}) to optimize traffic, energy efficiency, and reliability. However, many real-world applications do not expose explicit task graphs and often exhibit dynamic, data-dependent behavior, making them incompatible with static modeling approaches.

To address this, many efforts have explored dynamic task management as a way to implicitly support task graph construction. For instance, \cite{b18} applies benchmark-specific heuristics, while \cite{b19} introduces a code-centric preprocessing framework to automatically insert task creation and synchronization logic. While these methods provide runtime flexibility and offer some insight into task relationships, they fail to produce explicit, analyzable task graphs and are often tightly coupled to specific workloads. Other approaches, such as \cite{b20}, construct directed acyclic graphs (DAGs) for scheduling purposes, but these are typically fixed to specific task models and lack general-purpose analytical value.

Most importantly, existing methods fail to incorporate cache coherence interactions, particularly those arising from shared-data accesses that trigger coherence events such as invalidations, updates, or cache-to-cache transfers. These interactions are critical in shared-memory systems, where tasks accessing overlapping data can generate significant coherence traffic—even without explicit producer–consumer relationships. Conventional models \cite{b20,b21} tend to focus solely on direct dataflow relationships, capturing only explicit communication while overlooking coherence-induced delays, and implicit dependencies that significantly shape runtime behavior. As a result, designs based on such task graphs often lead to the misalignment between design-time assumptions and runtime performance, limiting the effectiveness of downstream decisions—such as task mapping, scheduling, or NoC optimization—and ultimately results in suboptimal system-level outcomes.

Coherence-aware task graph modeling for realistic applications is therefore essential. By explicitly capturing inter-task dependencies induced by coherence behavior, such models offer a more faithful representation of runtime dynamics in shared-memory systems. This enhanced accuracy provides a stronger foundation for system-level analysis and architectural optimization. It not only bridges the gap between real-world workloads and existing research methodologies but also empowers more informed and effective design strategies—including, but not limited to, NoC-level design.

In this paper, we highlight three key observations:

\begin{itemize}
    \item Many real-world applications lack explicit task graphs and exhibit dynamic, data-dependent behavior, making them incompatible with current design strategies that rely on predefined, statically defined graphs.

    \item Although some approaches targeting realistic applications enable flexible task management through dynamic runtime techniques, they typically do not produce analyzable task graphs and often rely on application-specific features. Even when task graphs are constructed, they are usually designed for scheduling purposes rather than general-purpose system analysis, which limits their broader applicability.

    \item Most critically, cache coherence interactions are largely overlooked in existing models, despite their substantial impact on inter-task communication patterns and system-level performance in shared-memory architectures.
\end{itemize} 
To address these gaps, we propose CoTAM, a Coherence-Aware Task Graph Modeling framework that explicitly captures coherence-induced inter-task dependencies. CoTAM provides a unified and analyzable model that supports realistic workloads with diverse characteristics, bridging the gap between dynamic application behavior and system-level design decisions.

Our detailed contributions are as follows: (i) We develop a novel model that decouples coherence behavior from conventional computation and communication time, enabling a more precise and interpretable analysis of coherence impact. (ii) We introduce a novel framework that infers inter-task dependencies based on runtime coherence behavior. This framework learns metric weights and computes directional dependency scores that capture asymmetric coherence-induced interactions between tasks. (iii) Building on these directional scores, we develop a novel coherence-aware task graph construction framework by classifying edges into varying types of dependencies, which enforces acyclicity through cycle elimination, yielding a directed acyclic graph (DAG) suitable for scheduling and analysis. We evaluate our proposed framework, CoTAM, using the Gem5 simulator with PARSEC benchmarks. Results demonstrate that CoTAM consistently outperforms existing strategies, leading to improved system performance and efficiency.

\section{Related Work}
\label{sec: related work}
In recent years, task graph modeling has gained increasing attention due to its foundational role in system-level analysis and design. In particular, it supports decisions in domains such as Network-on-Chip (NoC) design, where task graphs inform strategies like task mapping and mapping-routing co-optimization. For example, EagerMap \cite{b2} use hierarchical greedy grouping to minimize inter-task communication, while \cite{b3} uses mixed-integer linear programming with heuristics to jointly optimize energy, and network contention. MARCO \cite{b9} further combines tabu search and reinforcement learning for co-optimization, 
while LAMP \cite{b23} introduces a software-hardware co-design that uses reinforcement learning and load-balancing strategies for efficient multipath transmission.

However, these techniques assume the availability of a static, defined task graph, which is often unrealistic for modern applications. Real-world workloads frequently involve implicit, data-driven, and dynamically evolving dependencies that are difficult to model statically. To address this, several efforts provide the potential for task graph generation. For example, \cite{b18} tailors task graph construction to individual benchmarks, using coarse-grain random task spawning for some workloads and statically encoding data dependencies for pipelined applications. Similarly, \cite{b19} presents a code-centric preprocessing framework that automatically injects task creation and synchronization calls, supporting configurable loop-level parallelism through manually specified grain size and task count. While these approaches rely on runtime behavior and do not explicitly construct analyzable graphs, other methods—such as \cite{b20}—do explicitly construct a directed acyclic graph (DAG) to represent task dependencies. It further proposes a mixed-integer linear programming (MILP) formulation with a list-based scheduling algorithm for non-preemptive, data-dependent periodic tasks.  However, such DAGs are typically tailored to fixed task models or scheduling objectives and are not intended for general-purpose analysis across diverse, realistic workloads.


Moreover, these existing approaches do not account for cache coherence interactions, which play a critical role in shaping inter-task communication. This omission leads to task graphs that fail to reflect actual runtime behavior, ultimately limiting their effectiveness in guiding system-level design decisions. 

There is, therefore, an urgent need for coherence-aware task graph modeling that supports realistic applications and bridges the gap between application behavior and architectural optimization.

\section{Methodology}
\label{method}
\begin{figure}[htbp]
    \centering
        \includegraphics[width=1.01\linewidth]{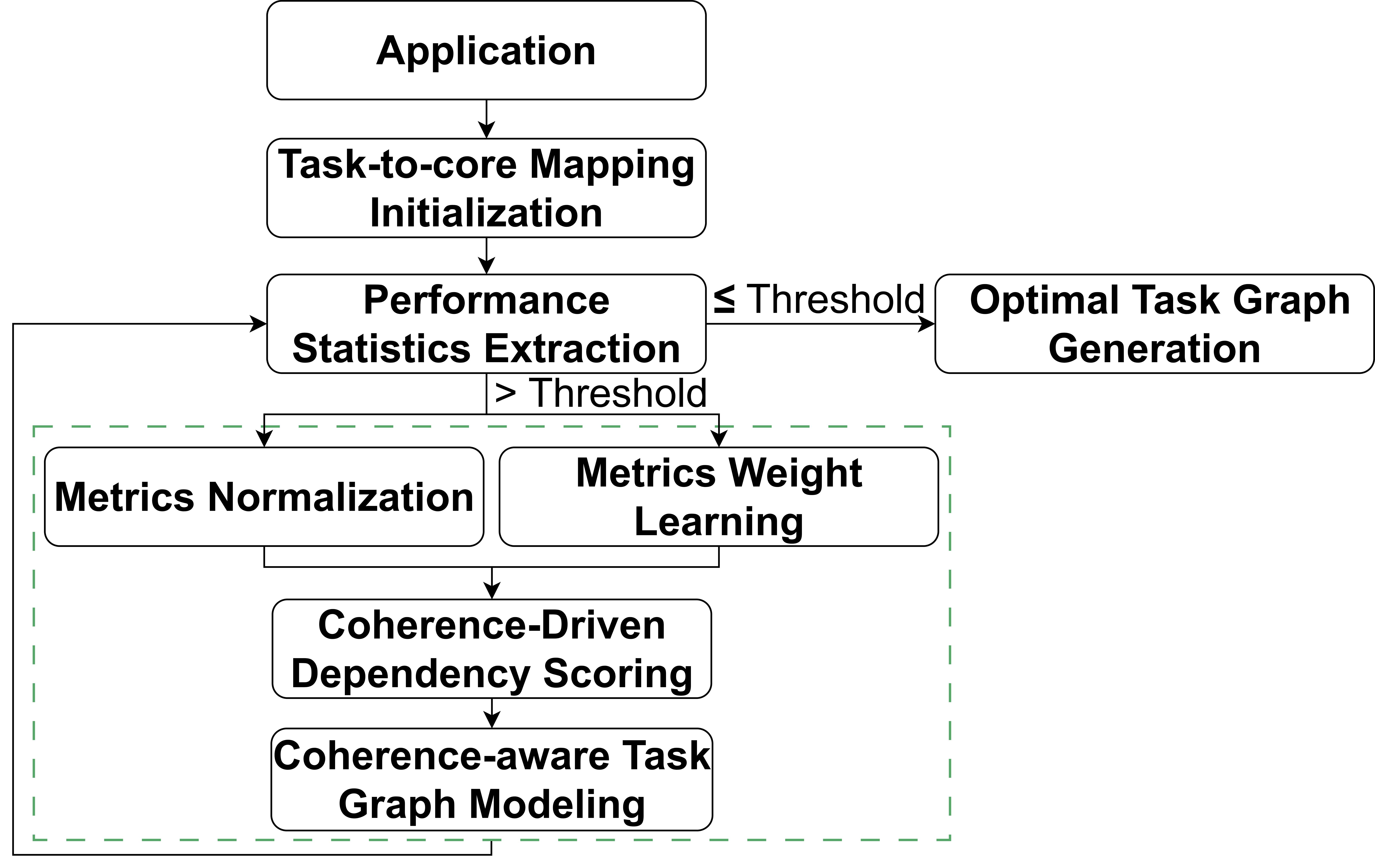}

    \caption{Workflow of CoTAM}
    \label{fig:workflow}
\end{figure}

In this section, we present CoTAM, our proposed Coherence-Aware Task Graph Modeling framework, as illustrated in Figure~\ref{fig:workflow}. CoTAM dynamically infers inter-task dependencies by analyzing runtime coherence metrics, decoupling coherence behavior from task execution patterns, and learning asymmetric, coherence-induced relationships that extend beyond traditional explicit communication. By capturing both direct and implicit coherence-induced interactions, CoTAM enables more accurate and realistic task graph modeling for real-world applications.

In addition to describing the core modeling pipeline, this section also introduces the adaptability and integration features of CoTAM, demonstrating its flexibility across a wide range of application scenarios and simulation environments.

 \subsection{Motivation for Coherence-Aware Task Graph Modeling}
 \label{motivation_coherence}

Cache coherence protocols maintain memory consistency by coordinating access to shared data through operations such as invalidations, updates, and cache-to-cache transfers. These coherence messages introduce additional latency, bandwidth overhead, and contention within the memory hierarchy and interconnect. More importantly, they expose implicit inter-task dependencies arising from shared data accesses—even in the absence of explicit producer–consumer relationships.

This coherence-induced behavior contrasts with what conventional task graphs typically capture, which are limited to explicit communication patterns—such as direct data flow from one task to another. Coherence events reveal a richer and more nuanced set of runtime interactions that go beyond these explicit dependencies. For instance, when two tasks frequently access the same cache line—whether for reads or writes—coherence actions like invalidations or updates are triggered. These interactions introduce implicit dependencies tied to the underlying coordination of shared memory resources, even when no direct communication exists between the tasks. Because traditional task graphs overlook these coherence-induced interactions, they provide an incomplete view of inter-task communication, ultimately underestimating execution overheads and misrepresenting runtime behavior in shared-memory systems.



Moreover, while many real-world applications do not expose predefined task graphs—on which existing design strategies often rely—they are still subject to the overhead imposed by coherence protocols. As a result, coherence-induced interactions remain an unavoidable source of execution cost. 

Incorporating coherence-induced dependencies into task graph modeling is therefore essential. It enables a more faithful representation of inter-task communication by capturing both explicit and implicit interactions. This, in turn, supports more accurate performance modeling and informs a wide range of system-level design decisions—including, for example, routing, scheduling, and resource allocation in Network-on-Chip (NoC) architectures.

\subsection{Coherence Behaviour Decoupling Model}
\label{motivation_graph}

As discussed in Section~\ref{introduction}, conventional task graph modeling overlooks coherence effects by focusing on explicit communication or dataflow dependencies. This omission makes it challenging to capture the subtle yet significant impact of coherence-induced interactions—such as invalidations, updates, and shared-data accesses—on inter-task behavior. To address this gap, our methodology explicitly isolates coherence behavior from traditional execution components, namely computation and communication. By disentangling coherence overhead, we enable a more precise attribution of its performance impact and facilitate the construction of task graphs for realistic applications that reflect implicit, coherence-aware relationships. This enhanced modeling approach provides a more faithful representation of runtime dynamics in shared-memory systems and lays a stronger foundation for downstream optimization in system-level design.

To gain insight into how coherence behavior affects execution, we introduce two key metrics that capture its indirect impact on computation and communication time, respectively:

\begin{itemize}
    \item Private cache hit ratio. This metric indirectly reflects the coherence overhead on computation. A higher hit ratio indicates that more data is served locally, thereby reducing coherence misses and the associated stall cycles. Conceptually, this relationship can be approximated as:
    \begin{equation}
        T_{\text{comp}} = \alpha \cdot \frac{1}{H_{\text{priv}}}
    \end{equation}
    where $T_{\text{comp}}$ denotes computation time, $H_{\text{priv}}$ is the private cache hit ratio, and $\alpha$ is a scaling factor that captures the sensitivity of computation time to hit ratio.
    
    \item Coherence‑induced message transmissions. This metric captures the impact of coherence on communication time. It reflects the volume of coherence-related traffic—including invalidations, updates, and other protocol messages—that contributes solely to communication time. Local coherence delays such as transient states and stalls are not included. We model this as:
    
    \begin{equation}
        T_{\text{comm}} = \beta \cdot M_{\text{cc}}
    \end{equation}
    where $T_{\text{comm}}$ is the communication time, $M_{\text{cc}}$ is the number of coherence messages, and $\beta$ is a proportionality constant translating message count into time overhead.

\end{itemize}

Beyond the decomposition into computation and communication time, we further define coherence time as a comprehensive measure that spans both domains. This metric quantifies the total delay introduced by coherence operations, including message transmission latency, write-backs, data fetches, invalidation handling, and state transitions. It provides a holistic view of coherence overhead and is modeled as:
\begin{equation}
        T_{\text{cc}} = T_{\text{comp}}^{\text{cc}} + T_{\text{comm}}^{\text{cc}}
\end{equation}
More explicitly:
\begin{equation}
        T_{\text{cc}} = \gamma_1 \cdot \left( \frac{1}{H_{\text{priv}}} \right) + \gamma_2 \cdot M_{\text{cc}}
\end{equation}
where $T_{\text{cc}}$ is the coherence time, and $\gamma_1$, $\gamma_2$ represent scaling factors for coherence-induced delays in computation and communication respectively.

Combining the above, we derive a layered execution time model:
\begin{equation}
    T_{\text{exec}} = \lambda_1 \cdot T_{\text{cc}} + \lambda_2
\end{equation}
where $T_{\text{exec}}$  denotes the execution time, incorporating both computation and communication components, $\lambda_1$ reflects the impact of coherence time on task execution, and $\lambda_2$ accounts for overheads unrelated to coherence.

This explicit decoupling forms the foundation of CoTAM’s design. Although the derived expressions are not directly embedded in CoTAM’s implementation, they offer valuable analytical insight into coherence effects that would otherwise be obscured within the overall task execution.

\subsection{Coherence Feature Extraction}
\label{extraction}

Although coherence mechanisms reflect how tasks interact through shared data, they are implemented at the core level in hardware. As a result, coherence overhead arises from inter-task data dependencies but manifests as core-level events during execution. This disconnect makes it challenging to directly observe coherence performance between tasks. Instead, we capture coherence-related metrics at the core level and later project them to the task level, as discussed in subsequent sections.

To support this, our methodology utilizes CCTA—the coherence analysis tool introduced in \cite{b17}—to collect detailed runtime coherence statistics, as summarized in Table~\ref{tab:coherence_measurement}, as well as system statistics as summarized in Table~\ref{tab:noc_measurement}. Although our goal is to build a coherence‐aware task graph, coherence events do not occur in isolation. The true cost of data sharing is reflected in system‐level performance: coherence‐induced message transmission and private cache hit ratio directly influence total execution time. By incorporating system metrics—such as overall execution time and stall‐free computation time—we weight each dependency not only by its raw coherence volume but also by its tangible impact on application performance. Consequently, blending coherence with system metrics yields a dependency graph that more accurately captures the relationships critical for end‐to‐end optimization.

\begin{table}[h!]
\centering
\caption{Coherence Behaviour Measurement}
\renewcommand{\arraystretch}{1.30} 
\begin{tabular}{|p{3.0cm}|p{4.8cm}|}  
\hline
\textbf{Feature of each core} & \textbf{Description} \\
\hline
CPU delay &  One type of coherence time—the time associated with write-hit operations in the private cache when in state S—is entirely caused by cache coherence protocol.\\
\hline
Write miss time &  One of coherence time, time of write miss behaviour caused by protocol as well as prefetching, write back, etc.\\
\hline
Read miss time &  One of coherence time, time of read miss behaviour caused by protocol as well as prefetching, write back, etc.\\
\hline
Coherence time & Total delay introduced by coherence operations, including the above three types of time. \\
\hline
Num\_l1\_l1\_message & The number of coherence-induced message transmission that a core sends to other cores' private caches. \\
\hline
Private cache hit ratio & The percentage of memory accesses by a core that are successfully served from its own private cache, without needing to fetch data from shared caches, other cores, or main memory. \\
\hline
\end{tabular}
\label{tab:coherence_measurement}
\end{table}

\subsection{Coherence-induced Metric Weighting Scheme}
\label{weight_learning}
To quantify the relative importance of each coherence-related metric in characterizing inter-task behavior, we compute the variance of each metric across all cores and apply a softmax transformation to derive a normalized weight distribution. The underlying rationale is that metrics with greater variability across cores are more discriminative, as they capture nuanced differences in coherence behavior that are likely to impact task dependencies. In contrast, metrics with little variation offer limited insight into coherence-induced performance asymmetries.

Formally, let \( \mathcal{M} = \{m_1, m_2, \dots, m_K\} \) denote the set of \( K \) coherence-related metrics (e.g., coherence time, execution time, cache hit ratio), and let \( \sigma^2(m_k) \) represent the variance of metric \( m_k \) across all cores. The corresponding weight \( w_k \) assigned to \( m_k \) is computed using a softmax-over-variance formulation:

\begin{equation}
w_k = \frac{\exp\left( \sigma^2(m_k) \right)}{\sum_{j=1}^{K} \exp\left( \sigma^2(m_j) \right)}
\end{equation}

This formulation yields a normalized weight distribution over \( K \) metrics, denoted as \( \{w_k\}_{k=1}^{K} \), where each weight reflects the relative informativeness of its corresponding metric. By ensuring that the weights sum to one, the model preserves interpretability and supports direct comparisons across metrics. These learned weights are then used to compute coherence-aware directional scores between task pairs (Section~\ref{scoring}), forming the foundation for fine-grained modeling of coherence-induced dependencies within the CoTAM framework.

\begin{table}[h!]
\centering
\caption{System Behaviour Measurement}
\renewcommand{\arraystretch}{1.30} 
\begin{tabular}{|p{3.0cm}|p{4.8cm}|}  
\hline
\textbf{Feature of each core} & \textbf{Description} \\
\hline
Busy time &  Time of core actively processing instructions or handling operations, not including waiting time for data. \\
\hline
Total execution time &   Time of completing all the tasks.\\
\hline
\end{tabular}
\label{tab:noc_measurement}
\end{table}

\subsection {Coherence-induced Dependency Scoring}
\label{scoring}

To infer coherence-induced dependencies between tasks, we propose the Coherence-induced Dependency Scoring (CDS), which quantifies the likelihood that one task (e.g., \( t_i \)) depends on another (e.g., \( t_j \)) based on runtime-observed coherence behavior. Unlike conventional dependency models that rely on predefined communication patterns, CDS dynamically captures emergent inter-task relationships driven by shared-memory access and coherence events. This enables fine-grained modeling of data interactions that are critical for understanding performance in shared-memory multicore systems.

\textbf{Metric Normalization.} 
Given a set of core-level metrics extracted in Section~\ref{extraction}—such as coherence time, execution time, and cache hit ratio—each metric \( m_k \) is normalized across all cores to ensure consistency across heterogeneous measurements. We apply max normalization, which scales each value relative to the maximum observed across all cores:

\begin{equation}
\hat{m}_k(c) = \frac{m_k(c)}{\max\limits_{c' \in \mathcal{C}} m_k(c')}
\end{equation}

Here, \( m_k(c) \) denotes the raw value of metric \( m_k \) on core \( c \), and \( \mathcal{C} \) represents the set of all cores. This normalization procedure consistently scales each metric to the range \([0, 1]\), preserving relative performance differences across cores while eliminating distortions caused by varying units or magnitudes. As a result, the normalized metrics \( \hat{m}_k(c) \) become directly comparable and can be effectively combined with the learned weights (Section~\ref{weight_learning}) for downstream dependency scoring.

\textbf{Directional Score Definition.}
Given a task-to-core mapping where task \( t_i \) is assigned to core \( c_i \) and task \( t_j \) to core \( c_j \), we define the \textit{directional coherence score} from \( t_i \) to \( t_j \), denoted as \( S(t_i, t_j) \), as:

\begin{equation}
S(t_i, t_j) = \sum_{k=1}^{K} w_k \cdot \hat{m}_k(c_j)
\end{equation}

Here, \( \hat{m}_k(c_j) \) represents the normalized value of coherence-related metric \( m_k \) observed on core \( c_j \), and \( w_k \) is its corresponding importance weight, learned using the softmax-over-variance approach described earlier.

It is important to note that the task-to-core mapping is assumed to be pre-existing—typically derived from a prior iteration or external strategy—and is used solely as a reference to assess directional coherence influence. As discussed in Section~\ref{extraction}, coherence-related metrics are captured at the core level rather than per task, making the task-to-core mapping essential for projecting these metrics onto individual tasks. This scoring formulation quantifies how the coherence behavior of the core executing task \( t_j \) may impact task \( t_i \), enabling the framework to infer asymmetric, coherence-induced dependencies between tasks.

\subsection{Coherence-aware Task Graph Modeling}
\label{modeling}

Based on the computed directional coherence scores, we construct a directed, coherence-aware task graph \( G = (\mathcal{T}, \mathcal{E}) \), where each node \( t_i \in \mathcal{T} \) represents a task, and a directed edge \( t_i \rightarrow t_j \in \mathcal{E} \) is added if the directional score \( S(t_i, t_j) \) exceeds a predefined threshold. This graph models inter-task dependencies that arise from coherence-induced interactions during execution.

To systematically characterize the strength of these dependencies, we introduce two threshold parameters: a \textit{strong coherence threshold} \( T_{\text{strong}} \) and a \textit{weak coherence threshold} \( T_{\text{weak}} \), such that \( T_{\text{strong}} > T_{\text{weak}} \). These thresholds allow us to classify each dependency score into one of three categories:

\begin{itemize}
    \item \( S(t_i, t_j) > T_{\text{strong}} \): Strongly coherence-dependent
    \item \( T_{\text{weak}} < S(t_i, t_j) \leq T_{\text{strong}} \): Moderately coherence-dependent
    \item \( S(t_i, t_j) \leq T_{\text{weak}} \): Weakly coherence-dependent (excluded from the graph)
\end{itemize}

Only strong and moderate dependencies are retained as directed edges in the final graph. An edge \( t_i \rightarrow t_j \) indicates that task \( t_i \) exhibits coherence-induced dependence on task \( t_j \), implying potential delays or synchronization penalties due to shared data access patterns. Since the directional score \( S(t_i, t_j) \) is inherently asymmetric—i.e., \( S(t_i, t_j) \neq S(t_j, t_i) \)—the resulting task graph captures non-reciprocal, runtime-specific coherence behavior and is asymmetric by design.

To ensure the correctness of the graph and its suitability for further analysis and system design, we enforce acyclicity by iteratively detecting and removing edges that form cycles:

\begin{equation}
\text{while } G \notin \text{DAG:} \quad \text{remove edge } (t_u, t_v) \in \text{cycle}(G)
\end{equation}

This results in a refined dependency graph \( G' \), which is a Directed Acyclic Graph (DAG). In this graph, each node corresponds to a task, and each directed edge \( t_i \rightarrow t_j \) indicates that task \( t_i \) exhibits a coherence dependency on task \( t_j \), as inferred from their associated core-level metrics. Each edge is further weighted to reflect the strength of the dependency, based on a weighted combination of normalized coherence and system-level behaviors, as detailed in Section~\ref{scoring}. The acyclic structure enables coherence-aware parallel execution and supports system-level modeling of task behavior.

\subsection{Adaptability and Integration of CoTAM}
To ensure practical applicability, the proposed Coherence-Aware Task Graph Modeling (CoTAM) framework is designed to accommodate a wide range of workloads, mapping strategies, coherence protocols, and system configurations. This section outlines the supported components, input/output formats, integration flexibility, and the current scope of CoTAM, offering a clear perspective on its extensibility and applicability.

\textbf{Supported Configurations.}
CoTAM supports a wide range of task mapping and routing strategies, cache coherence protocols, and architectural topologies. This inherent sensitivity underscores the importance of supporting diverse configurations to ensure accurate and representative task graph modeling. CoTAM’s flexibility enables its application to realistic multicore systems without requiring any modifications to the application logic or the underlying simulation infrastructure. The framework is designed for seamless integration into existing workflows and is adaptable to a variety of system configurations. Table~\ref{tab:supported_configs} summarizes the currently supported features and configurations.

\begin{table}[h]
\centering
\caption{Supported Features and Configurations in CoTAM}
\label{tab:supported_configs}
\renewcommand{\arraystretch}{1.30}
\begin{tabular}{|p{3.8cm}|p{4.0cm}|}
\hline
\textbf{Category} & \textbf{Supported Options} \\
\hline
Task mapping and routing strategies & All strategies that can be used in the iterative loop, e.g., heuristic-based mapping or Reinforcement Learning-based routing \\
\hline
Coherence protocols & Directory-based MESI and extended protocol families derived from it, including CHI \\
\hline
System topologies & Mesh, Torus, Pt2Pt, Crossbar, Fat Tree, Flattened Butterfly \\
\hline
Application types & Workloads that can exhibit coherence behavior \\
\hline
Architecture & Architectures supporting cache coherence behavior, including NoC-based multicore systems \\
\hline
\end{tabular}
\end{table}

\textbf{Input and Output Format.}
CoTAM operates on runtime information collected from simulations, leveraging both coherence-related and system-level traces to extract and construct inter-task dependencies induced by coherence behavior. The resulting structure is modeled as a directed acyclic graph (DAG) that captures the coherence-induced interactions among tasks.

As summarized in Table~\ref{tab:CoTAM_io}, the input to CoTAM includes the application workload in the form of .rcS scripts, which configure and launch benchmarks in Gem5 full-system mode. Additionally, for each simulation iteration, performance metrics extracted from stats.txt—Gem5’s default output file, extended with coherence metrics by CCTA \cite{b17}—serve a dual role: they are recorded as output and fed back into the framework as input for iterative modeling and dependency refinement.

On the output side, CoTAM generates a coherence-aware task graph, represented as a directed acyclic graph (DAG) in JSON format for downstream use. For realistic applications that exhibit similar coherence characteristics, this output can be directly integrated into downstream design flows, such as task mapping and scheduling optimization—facilitating tighter coupling between runtime coherence behavior and architectural decision-making.

\begin{table}[h]
\centering
\caption{Input and Output Description of CoTAM}
\label{tab:CoTAM_io}
\renewcommand{\arraystretch}{1.30}
\begin{tabular}{|p{3.5cm}|p{4.3cm}|}
\hline
\textbf{Type} & \textbf{Description} \\
\hline
Input & 
(i) .rcS scripts for application. \newline
(ii) stats.txt for system and coherence metrics. \\
\hline
Output & 
(i) DAG in JSON format \newline
(ii) stats.txt for system and coherence metrics.  \\
\hline
\end{tabular}
\end{table}

\textbf{Applicability and Integration Scope.}
CoTAM is designed for broad applicability across realistic multicore simulation environments where cache coherence significantly impacts performance. It is particularly well-suited for systems employing shared-memory architectures and executing workloads with frequent data sharing, inter-task communication, or memory contention. Such scenarios commonly arise in domains like parallel scientific computing, media processing, and graph analytics—where coherence protocols influence both execution time and communication patterns.

The framework integrates seamlessly into simulation workflows, such as those based on gem5 with Ruby—without requiring any modifications to the application source code. CoTAM is agnostic to the scheduling model and can be applied in both exploratory research and practical performance evaluations. It supports both post-simulation analysis using offline traces and runtime integration for iterative modeling.

However, the current implementation depends on iterative simulation to capture and update coherence metrics. Consequently, runtime overhead may increase for very large-scale applications, especially those with long execution phases or complex sharing behaviors. As such, CoTAM is particularly suited for medium-scale workloads, where coherence effects are prominent but remain tractable within simulation constraints.

\section{Experiment}
The experiments are conducted using the Gem5 simulator \cite{b12}, which provides full-system simulation in a realistic cache-coherent environment. It supports directory-based MESI protocols—commonly adopted in both academic research and industry standards such as the Coherent Hub Interface (CHI)—enabling accurate modeling of coherence interactions. Table~\ref{tab:setup} summarizes the detailed system configuration. For evaluation, we select three representative workloads from the PARSEC benchmark suite \cite{b11}—Canneal, Blackscholes (Bla), and Fluidanimate (Flu)—as they represent distinct workload types, offering a comprehensive assessment of CoTAM’s performance. These benchmarks span a diverse range of computational and memory behaviors: Canneal is memory-intensive, Blackscholes is compute-bound, and Fluidanimate features fine-grained parallelism. This diversity makes them well-suited for evaluating the effectiveness of coherence-aware task graph modeling.

\begin{table}[h!]
\centering
\caption{Platform Parameters}
\renewcommand{\arraystretch}{1.30} 
\begin{tabular}{|p{3.5cm}|p{4.3cm}|} 
\hline
\textbf{Platform Parameters} & \textbf{Values} \\
\hline
System Architecture  & NoC-based multicore systems \\
\hline
Virtual channels per port & 4 \\
\hline
Flow control & Credit-based \\
\hline
Frequency & 2 GHz \\
\hline
Flit size & 128 bits \\
\hline
L1D Cache & Private, 64 KB \\
\hline
L2 Cache & Shared, 2 MB \\
\hline
Memory size & 512 MB \\
\hline
Cacheline size & 64 B \\
\hline
Cache coherence protocol & Dirctory-based MESI \\
\hline
Topology & 2D Mesh\\
\hline
Routing Strategy & XY Routing \\
\hline
\end{tabular}
\label{tab:setup}
\end{table}

To evaluate the advantages of our CoTAM, we compare it with a task parallelism strategy inspired by TP-PARSEC \cite{b19}. While TP-PARSEC is designed for execution on real hardware and does not explicitly construct task graphs, it represents a widely adopted approach for parallelizing applications through automatic task extraction via code preprocessing. In our evaluation, we implement a comparable strategy within the Gem5 simulation environment, adopting Pthread-style task parallelism without modeling inter-task dependencies. This allows us to capture the core idea of TP-PARSEC while adapting it to our simulated setup. Given the lack of existing methods that construct task graphs for realistic workloads, we use this TP-PARSEC–inspired method as a reference. To ensure fairness, both approaches adopt similar Pthread-like task structures, ensuring consistency in runtime behavior and parallel execution.


\subsection{CoTAM’s Enhancement of Speedup}

\begin{figure}[htbp]
    \centering
        \includegraphics[width=1.01\linewidth]{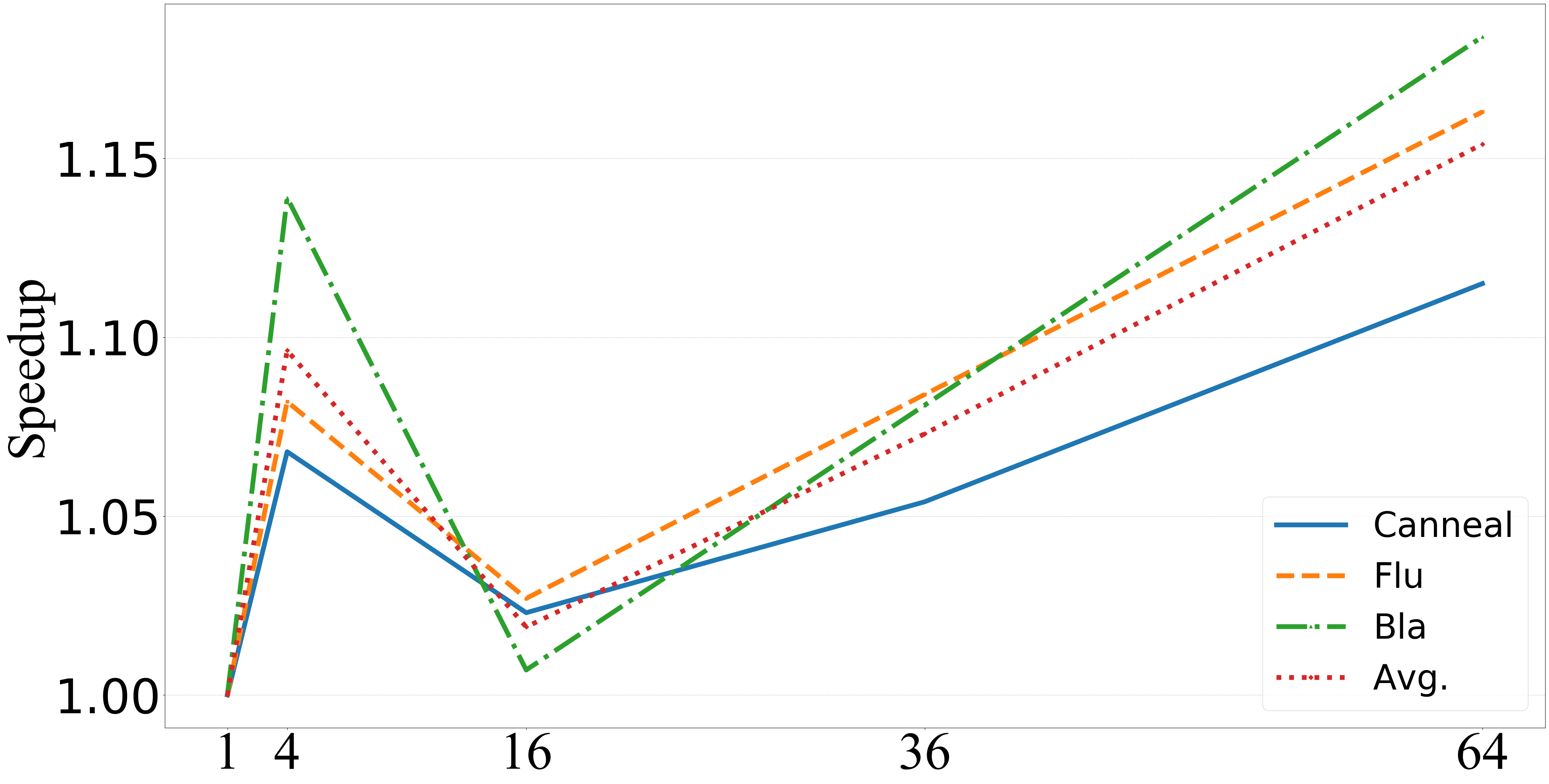}

    \caption{Speedup comparison across different workloads.}
    \label{fig:speedup_comparison}
\end{figure}

We compare the speedup achieved by our CoTAM framework against the TP-PARSEC-inspired task parallelism strategy across varying core counts (1, 4, 16, 36, and 64). As shown in Figure~\ref{fig:speedup_comparison}, CoTAM consistently delivers higher speedups across all configurations and workloads, with average improvements of $1.097\times$, $1.019\times$, $1.073\times$, and $1.154\times$, respectively.

This improvement stems from a fundamental difference in how each method handles task relationships. While TP-PARSEC automatically extracts tasks through code preprocessing, it does not construct an explicit task graph and therefore lacks global visibility into inter-task dependencies. Instead, it generates task creation strategies tailored to individual workloads, without modeling the broader structure of task interactions. In contrast, CoTAM constructs a coherence-aware task graph that captures implicit coherence-induced interactions observed at runtime. By exposing these dependencies, CoTAM provides a more informed view of task interactions, enabling better exploitation of parallelism and better system-level performance, highlighting the potential for further enhancement and refined design of coherence-aware task graph.

The variation in performance improvement across core counts can be attributed to how each method scales with increasing system complexity. At lower core counts (e.g., 4 cores), CoTAM effectively captures fine-grained coherence interactions by analyzing runtime behavior, enabling a more accurate representation of inter-task dependencies. As core counts increase (e.g., to 64 cores), coherence traffic and interference grow significantly, and CoTAM’s ability to isolate and quantify coherence effects becomes increasingly valuable. By decoupling coherence behavior from overall execution and inferring coherence-driven dependencies, CoTAM constructs task graphs that better reflect true runtime interactions. This deeper insight allows for more informed optimization and results in greater speedup as coherence-induced delays become a more dominant performance factor. The smaller improvement observed at 16 cores reflects a transitional point where workload parallelism is already largely exploited by TP-PARSEC, and coherence effects are not yet the primary bottleneck. Overall, this trend highlights CoTAM’s adaptability and its growing relevance in capturing coherence behavior as system scale increases.

\subsection{Analysis of CoTAM’s Impact Across Performance Dimensions}
In this section, we evaluate both NoC and cache coherence performance to better understand the impact of coherence dynamics on task graph generation across different types of workloads.

\begin{figure}[htbp]
    \centering
    
    \begin{subfigure}[b]{0.49\textwidth}
        \centering
        \includegraphics[width=\linewidth]{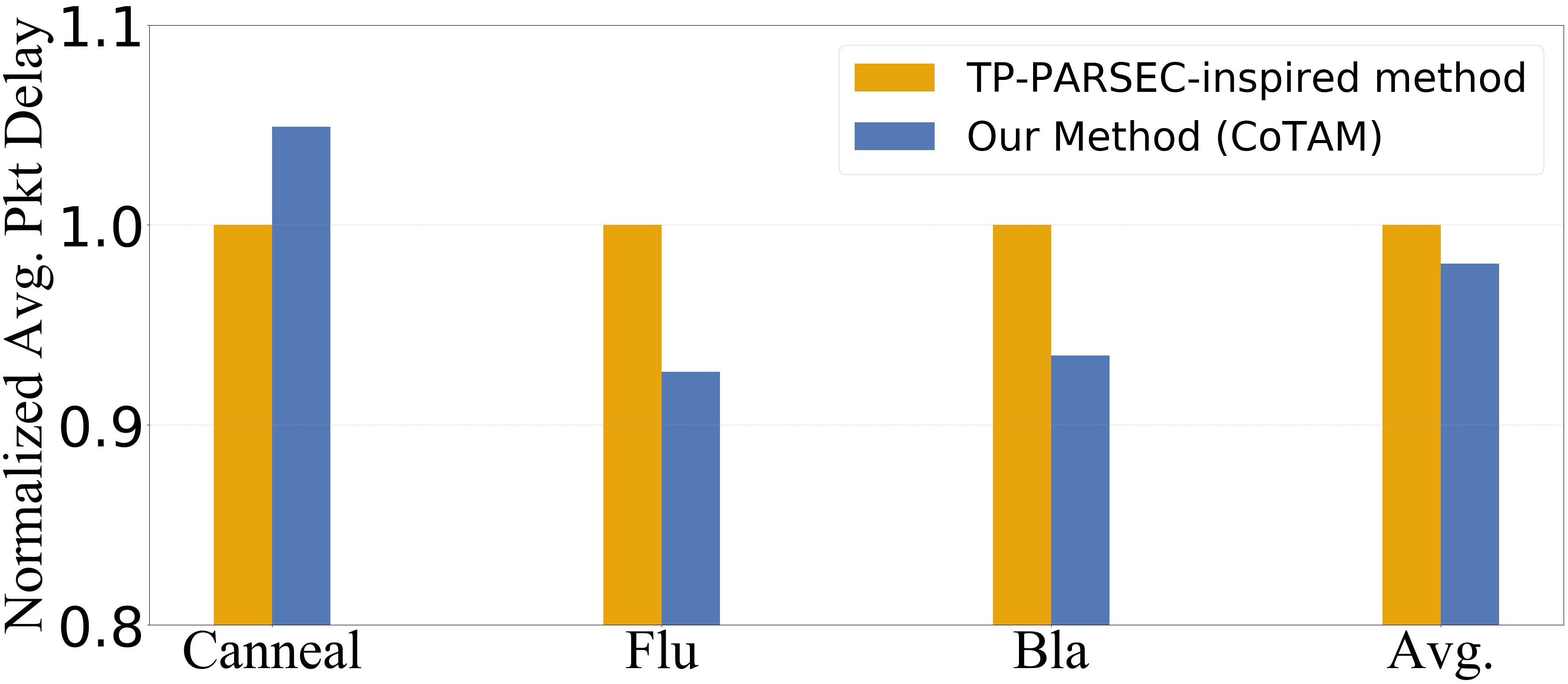}
        \caption{Average packet delay}
        \label{fig:delay_build}
    \end{subfigure}

    \begin{subfigure}[b]{0.49\textwidth}
        \centering
        \includegraphics[width=\linewidth]{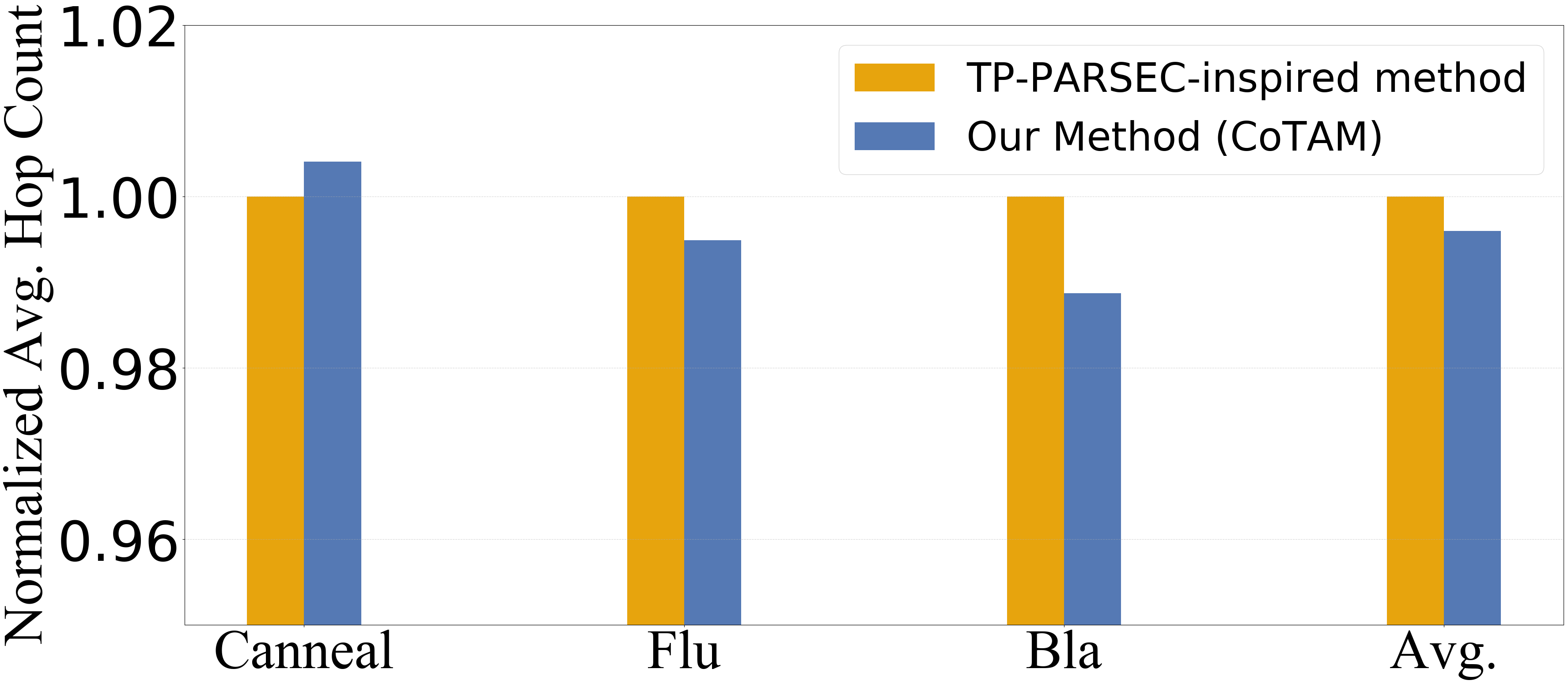}
        \caption{Average hop count}
        \label{fig:hop_build}
    \end{subfigure}

    \begin{subfigure}[b]{0.49\textwidth}
        \centering
        \includegraphics[width=\linewidth]{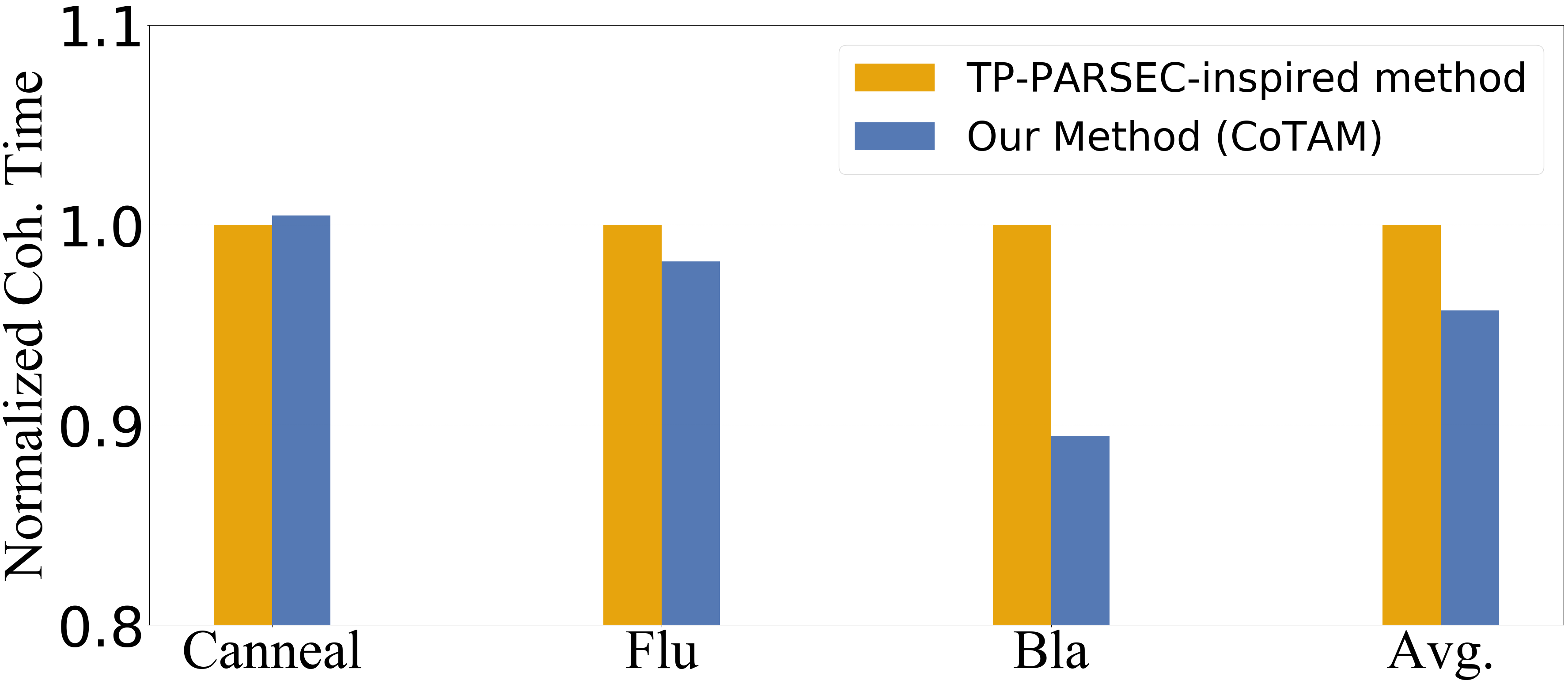}
        \caption{Total coherence time}
        \label{fig:cohtime_build}
    \end{subfigure}

    \caption{Normalized performance comparison between the two methods: (a) Average packet delay and (b) Average hop count, and (c) Total coherence time.}
    \label{fig:build}
\end{figure}

As shown in Figure~\ref{fig:build}, CoTAM achieves up to a 1.93\% reduction in average packet delay, a 0.40\% reduction in average hop count and a 4.28\% reduction in total coherence time compared to the TP-PARSEC–inspired task parallelism strategy.

Delving into the details, Figures~\ref{fig:build} reveal that memory-intensive workloads such as Canneal exhibit a different trend: CoTAM results in slightly higher average packet delay, average hop count and coherence time compared to the TP-PARSEC–inspired method. This is primarily due to Canneal’s frequent and large-scale shared-memory accesses, which generate dense coherence traffic and complex inter-task dependencies.

CoTAM captures these coherence-induced dependencies by preserving both strong and moderate relationships in its task graph. For Canneal, this results in a conservative DAG structure with reduced concurrency and elevated synchronization cost. These retained dependencies increase overall communication volume and contention in both the NoC and coherence systems, contributing not only to higher delay and coherence time but also to a slight increase in average hop count, as shown in Figure~\ref{fig:hop_build}. This longer routing distance reflects the distributed nature of coherence interactions under CoTAM’s detailed dependency model. Together, these factors explain the relatively smaller speedup achieved for Canneal in Figure~\ref{fig:speedup_comparison}.

In contrast, Flu and Bla follow the expected trend, showing improvements across all three metrics: delay, coherence time, and hop count. Flu, with its fine-grained parallelism and moderate data sharing, benefits from CoTAM’s accurate modeling of necessary dependencies without incurring excessive overhead. The resulting task graph preserves important synchronization while maintaining low communication cost, leading to improved execution efficiency. Similarly, Bla—being compute-bound with minimal shared-memory interaction—generates limited coherence activity. CoTAM’s task graph in this case introduces negligible communication overhead, enabling lower delay, shorter average routing paths, and reduced coherence time. These characteristics contribute to the more pronounced speedup seen in Figure~\ref{fig:speedup_comparison} for both Flu and Bla.

Overall, these findings highlight CoTAM’s adaptability across diverse workload characteristics. By explicitly capturing coherence-induced dependencies, CoTAM provides consistent performance improvements for applications with balanced or sparse coherence behavior, such as Flu and Bla, where it reduces coherence time, packet delay, and hop count. In more coherence-intensive workloads like Canneal, CoTAM faithfully models dense coherence interactions, which introduces additional overhead but preserves execution semantics and ensures analytical accuracy. Despite the modest speedup in such scenarios, CoTAM’s ability to generalize across workloads with varying communication and coherence profiles demonstrates its robustness as a coherence-aware task graph modeling framework—suitable for system-level analysis in realistic multicore environments.

\subsection{CoTAM's Tolerance on Task Granularity}



\begin{figure}[htbp]
    \centering
        \includegraphics[width=1.01\linewidth]{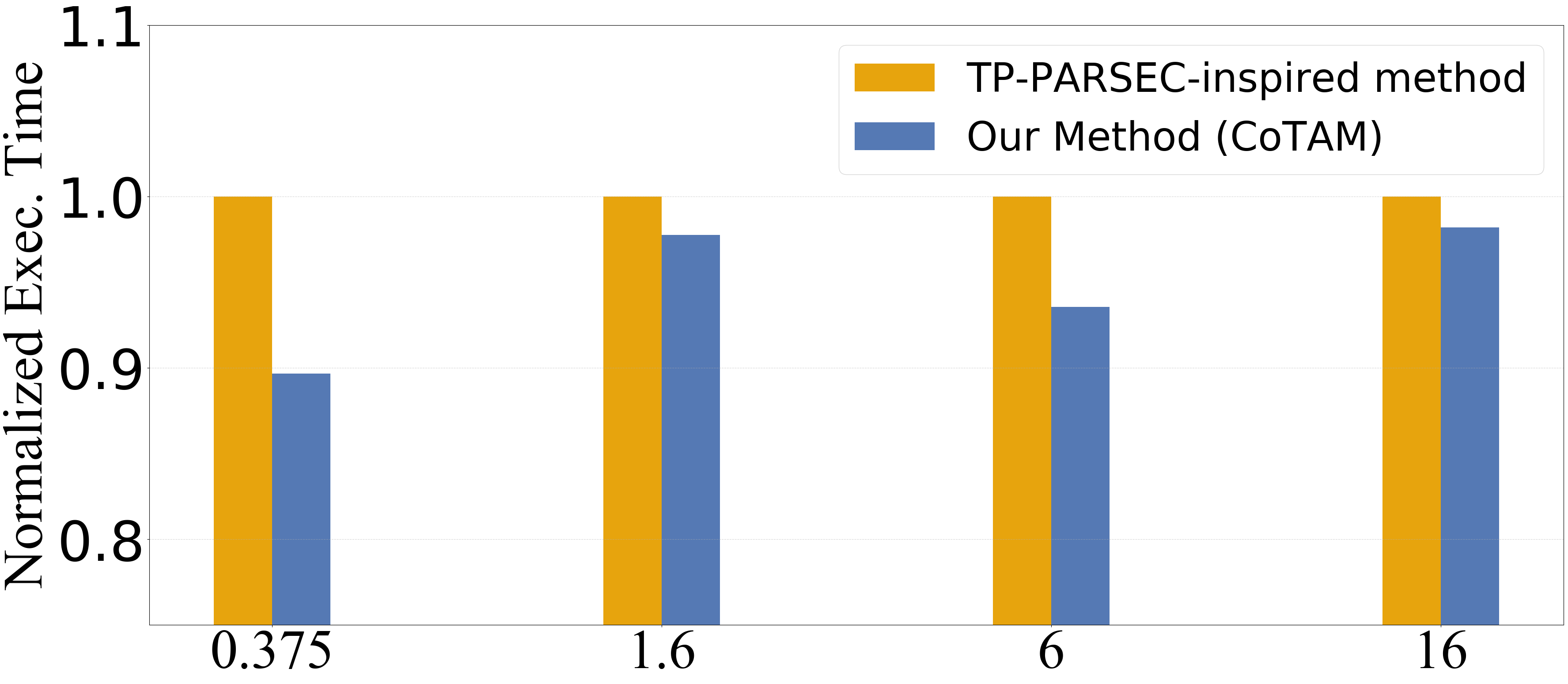}

    \caption{Comparison of execution times across varying task granularities.}
    \label{fig:task_granularity}
\end{figure}

In this section, we evaluate the impact of task granularity on CoTAM using the metric $T/C$, defined as the number of tasks divided by the number of cores. As shown in Figure~\ref{fig:task_granularity}, when $T/C = 0.375$, CoTAM achieves its largest execution time reduction of 11.5\% compared to the TP-PARSEC-inspired task parallelism strategy, indicating its strong effectiveness in coarse-grained scenarios. In such settings, coherence overhead tends to be more pronounced, and CoTAM’s ability to model coherence-induced dependencies enables it to expose critical performance bottlenecks and better reflect runtime behavior.

As task granularity increases—at $T/C = 1.6$ and $T/C = 6$—CoTAM continues to deliver notable execution time reductions of 2.3\% and 6.8\%, respectively. These improvements indicate that the framework remains effective in capturing meaningful coherence-induced interactions across a range of granularity levels.  However, when $T/C$ reaches 16, the benefit drops to 1.8\%, reflecting diminishing returns at extremely fine granularity. In this regime, coherence delays per task become less significant, coherence-induced dependencies are harder to distinguish from background execution noise, and performance is increasingly constrained by system-level overheads such as scheduling and synchronization rather than coherence behavior.

Based on these results, we recommend maintaining a $T/C$ ratio between 0.5 and 6 to balance coherence visibility and runtime efficiency, thereby achieving optimal performance under CoTAM.

\subsection{CoTAM's Tolerance on Coherence Threshold}
\begin{figure}[htbp]
    \centering
        \includegraphics[width=1.07\linewidth]{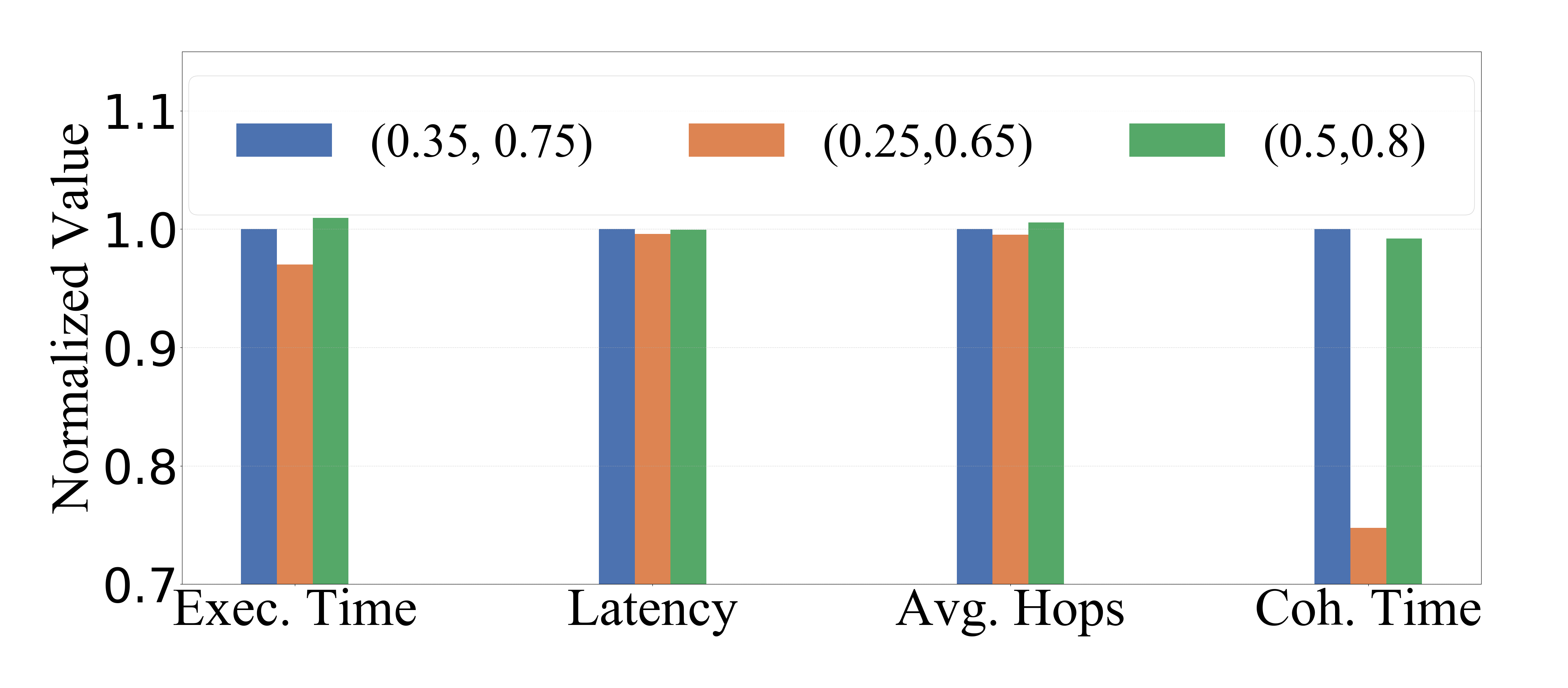}

    \caption{Comparison of different coherence threshold pairs.}
    \label{fig:threshold}
\end{figure}

To determine an effective coherence threshold pair for CoTAM’s dependency modeling, as introduced in Section~\ref{modeling}, we evaluate three configurations: a moderate sensitivity threshold $(0.35, 0.75)$ serving as the reference point, a relaxed sensitivity threshold $(0.25, 0.65)$ that admits more weak coherence-induced dependencies, and a strict sensitivity threshold $(0.50, 0.80)$ that captures only the strongest interactions. As shown in Figure~\ref{fig:threshold}, the relaxed configuration delivers the best overall performance—reducing execution time by 2.98\% and 3.91\%, average packet latency by 0.39\% and 0.34\%, average hop count by 0.49\% and 1.05\%, and coherence time by 25.23\% and 24.71\%, compared to the moderate and strict configurations, respectively.

The improvement in average hop count—defined as the total number of hops traversed per delivered packet—reflects enhanced spatial locality and more efficient downstream decisions, e.g., routing. While NoC-level gains are relatively modest, the substantial reduction in coherence time under the relaxed threshold significantly contributes to lowering overall execution time.

These results suggest that the relaxed threshold pair $(0.25, 0.65)$ offers an optimal trade-off between capturing meaningful coherence-induced dependencies and maintaining manageable modeling complexity. It effectively filters out low-impact interactions while retaining those that materially influence performance, making it the recommended configuration for CoTAM’s coherence metrics analysis and task graph construction.

\subsection{CoTAM's Adaptability on NoC Design}
As discussed in Section~\ref{introduction}, task graph modeling directly influences NoC design, such as task mapping. In this section, we evaluate CoTAM’s adaptability in guiding mapping decisions. To ensure a fair and interpretable comparison, we adopt a simple heuristic-based mapping strategy, which serves as the foundation for many existing approaches, including those proposed in \cite{b2,b3}.

\begin{figure}[htbp]
    \centering
    
    \begin{subfigure}[b]{0.49\textwidth}
        \centering
        \includegraphics[width=\linewidth]{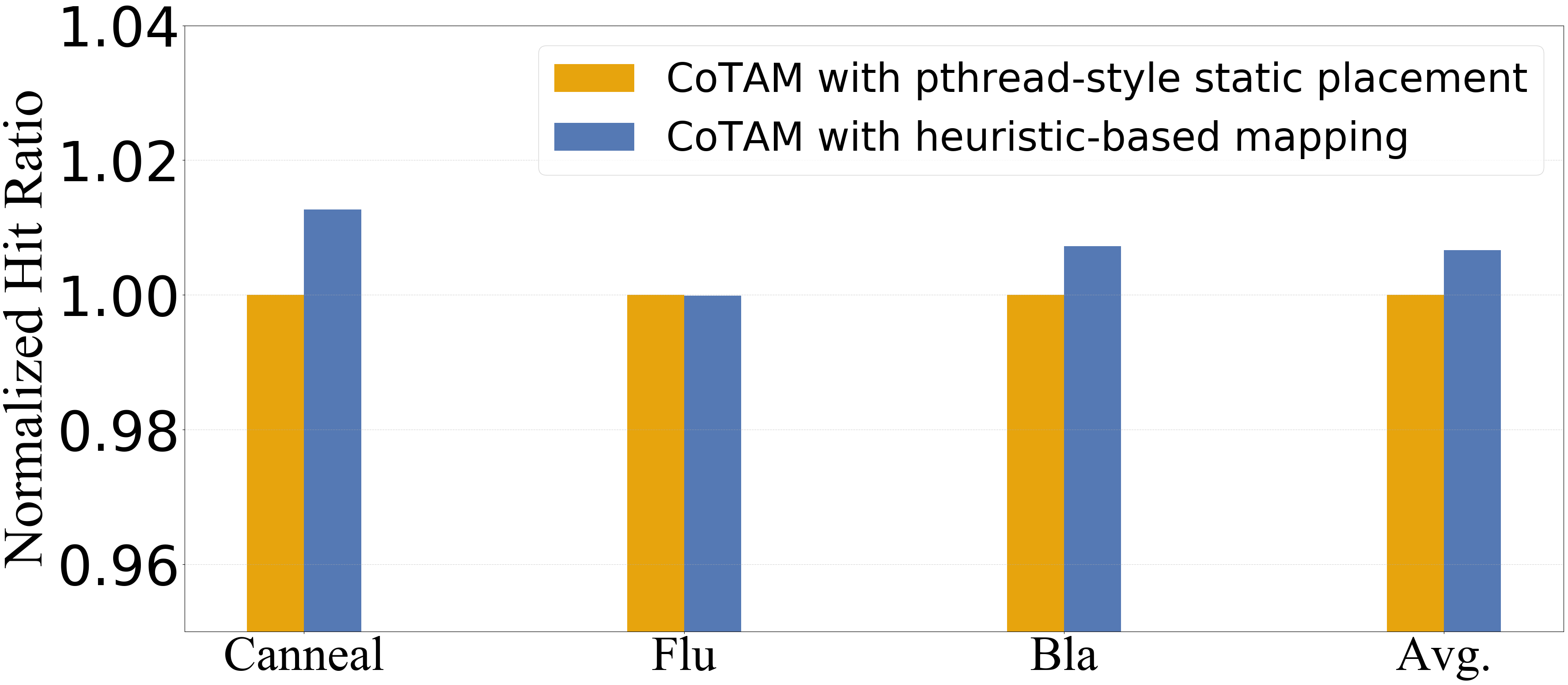}
        \caption{Private cache hit ratio}
        \label{fig:coherence}
    \end{subfigure}

    \begin{subfigure}[b]{0.49\textwidth}
        \centering
        \includegraphics[width=\linewidth]{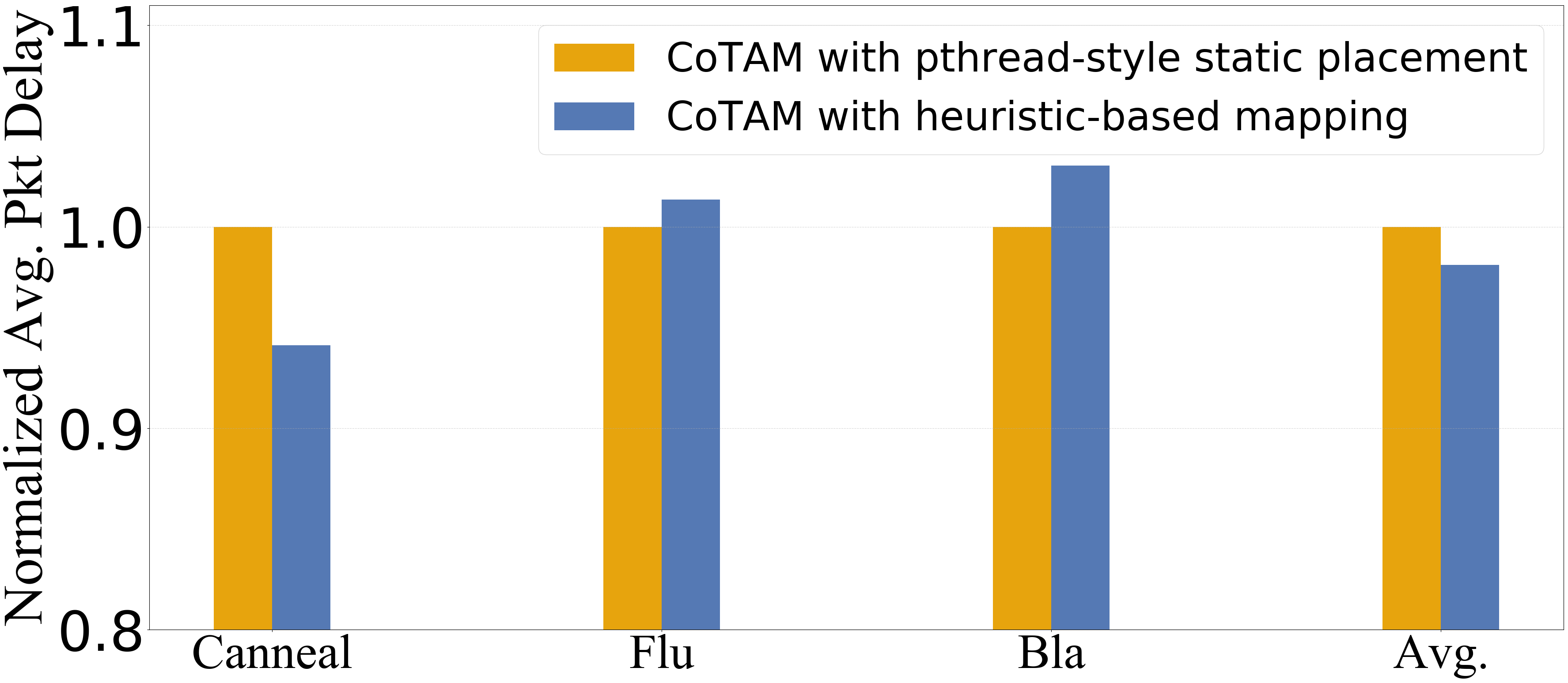}
        \caption{Average packet delay}
        \label{fig:delay}
    \end{subfigure}  

    \begin{subfigure}[b]{0.49\textwidth}
        \centering
        \includegraphics[width=\linewidth]{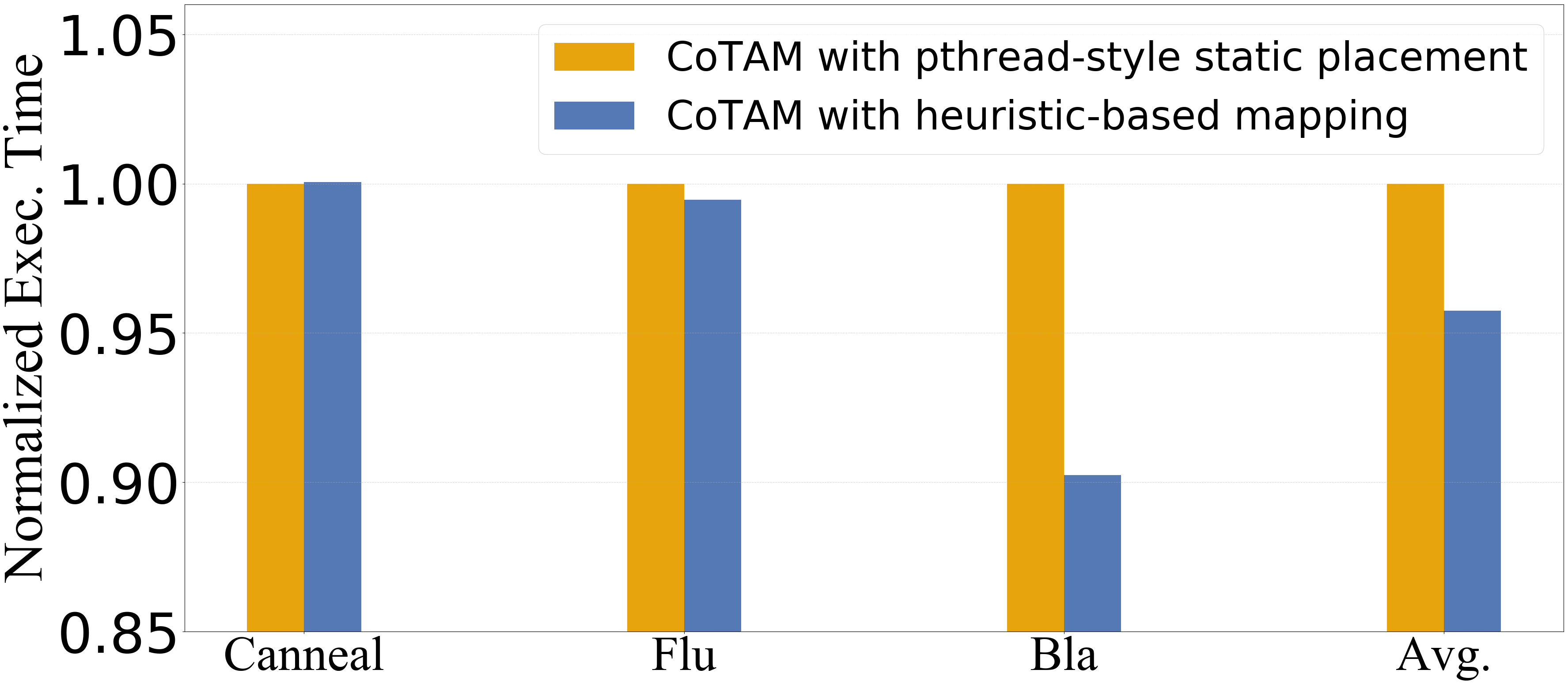}
        \caption{Execution time}
        \label{fig:exe_time}
    \end{subfigure}
      
    \caption{Normalized performance under two scenarios—CoTAM with pthread-style placement and CoTAM with heuristic-based mapping—measured by: (a) Private cache hit ratio, (b) Average packet delay, and (c) Execution time.}
    \label{fig:mapping}
\end{figure}

To evaluate this, we separately analyze cache coherence, NoC performance, and overall system performance under a Pthread-style static placement and a heuristic-based mapping strategy. As shown in Figure~\ref{fig:mapping}, CoTAM with heuristic mapping reduces execution time by 4.25\%, lowers average packet delay by 1.90\%, and increases private cache hit ratio by 0.6663\%, highlighting its ability to optimize both computation and communication aspects of task execution.

For cache coherence performance, CoTAM with heuristic-based mapping consistently improves private cache hit ratio across all benchmarks (Figure~\ref{fig:coherence}). By explicitly modeling inter-task coherence-induced dependencies, CoTAM reveals which tasks frequently access shared data. The heuristic mapping can then use this information to place coherence-interactive tasks closer together—often on the same or nearby cores—thereby reducing the likelihood of remote cache accesses and minimizing coherence-induced communication. Notably, benchmarks like Bla and Canneal exhibit over a 1\% increase in hit ratio, a significant improvement given that baseline ratios already exceed 96\%. While the absolute increase may appear small, even a 1\% gain at such high baseline levels translates to a meaningful reduction in coherence misses and memory access latency, contributing to system-level performance improvements.


In terms of NoC performance, CoTAM combined with heuristic-based task mapping leads to a reduction in average packet delay across benchmarks (Figure~\ref{fig:delay}), with an improvement of approximately 2\% compared to static Pthread-style placement. This improvement stems from CoTAM’s ability to expose coherence-induced dependencies, enabling the mapping strategy to co-locate or place interdependent tasks closer together in the NoC topology. As a result, coherence messages and data transfers traverse fewer hops, reducing network contention and improving inter-task communication efficiency. However, the overall reduction in packet delay is modest, as CoTAM is designed to model coherence behavior rather than directly optimize NoC traffic patterns. In benchmarks like Flu and Bla, where coherence traffic is balanced or limited, the benefit diminishes. In some cases, slight increases in packet delay suggest that when coherence-induced dependencies offer limited guidance, the heuristic mapping may inadvertently introduce suboptimal link utilization or localized congestion. These results indicate that while CoTAM enhances communication locality when coherence interactions are strong, its impact on NoC performance depends on the nature of coherence traffic in the workload.


For overall system performance (Figure~\ref{fig:exe_time}), CoTAM with heuristic-based mapping consistently matches or outperforms the Pthread-style approach. Notable gains are seen in Bla and Flu, where coherence-induced communication is more prominent. In these cases, CoTAM’s modeling of coherence dependencies enables more effective task placement, reducing execution stalls. In contrast, Canneal shows similar performance under both mappings, as its memory-bound nature limits the impact of coherence behavior. This aligns with the modest NoC delay reduction observed for Canneal, reinforcing that CoTAM is more beneficial in workloads where coherence overhead is a key performance factor.


In summary, CoTAM enhances coherence and NoC performance by capturing runtime inter-task dependencies induced by coherence through graph-based analysis. This results in a more accurate representation of system behavior and improved execution efficiency across diverse workloads. Its adaptability enables integration with existing design strategies—such as task mapping—even in realistic scenarios where task graphs are not explicitly available. These capabilities position CoTAM as a foundational bridge between dynamic real-world workloads and established system design methodologies.


\section{Conclusion}

This paper identifies three key challenges in task graph modeling for realistic applications: (i) many real-world workloads lack explicit task graphs or exhibit dynamic, data-dependent behavior that is difficult to capture statically, limiting the applicability of methods that rely on predefined structures; (ii) although some efforts targeting real-world applications explore dynamic runtime management, they typically do not produce analyzable task graphs and often depend on application-specific heuristics or scheduling-oriented representations, restricting general applicability; and (iii) cache coherence interactions—despite their significant impact on runtime behavior—are largely overlooked, leading to misalignment between design-time assumptions and runtime performance. 

To address these challenges, we propose CoTAM, a Coherence-Aware Task Graph Modeling framework that explicitly captures coherence-induced inter-task dependencies and supports general-purpose analysis across diverse workloads. Evaluated on PARSEC benchmarks using the Gem5 simulator, CoTAM achieves up to 15.4\% speedup over TP-PARSEC-inspired task parallelism, while enabling analysis across different task granularities and remaining compatible with existing multicore system designs. It not only bridges the gap between real-world workloads and existing research methodologies, but also underscores the importance of incorporating cache coherence into task graph modeling for realistic and effective system-level design.





\begin{thebibliography}{00}
\bibitem{b1} C. Wu et al., "A Multi-Objective Model Oriented Mapping Approach for NoC-based Computing Systems," in IEEE Transactions on Parallel and Distributed Systems, vol. 28, no. 3, pp. 662-676, 1 March 2017, doi: 10.1109/TPDS.2016.2589934. 

\bibitem{b2} E. H. M. Cruz, M. Diener, L. L. Pilla and P. O. A. Navaux, "An Efficient Algorithm for Communication-Based Task Mapping," 2015 23rd Euromicro International Conference on Parallel, Distributed, and Network-Based Processing, Turku, Finland, 2015, pp. 207-214, doi: 10.1109/PDP.2015.25.

\bibitem{b3} L. Mo, X. Li, A. Kritikakou and X. Zhai, "Contention and Reliability-Aware Energy Efficiency Task Mapping on NoC-Based MPSoCs," in IEEE Transactions on Reliability, vol. 74, no. 1, pp. 2010-2026, March 2025, doi: 10.1109/TR.2024.3377732.


\bibitem{b9}  Hui Chen, Zihao Zhang, Peng Chen, Xiangzhong Luo, Shiqing Li, and Weichen Liu. 2021. MARCO: A High-performance Task Mapping and Routing Co-optimization Framework for Point-to-Point NoC-based Heterogeneous Computing Systems. ACM Trans. Embed. Comput. Syst. 20, 5s, Article 54 (October 2021), 21 pages. https://doi-org.remotexs.ntu.edu.sg/10.1145/3476985.

\bibitem{b10} Vijay Nagarajan, Daniel J. Sorin, Mark D. Hill, David A. Wood, and Natalie Enright Jerger. 2020. A Primer on Memory Consistency and Cache Coherence (2nd. ed.). Morgan \& Claypool Publishers.

\bibitem{b11} ] C. Bienia, S. Kumar, J.P. Singh, K. Li, The PARSEC benchmark suite: Characterization and architectural implications, in: Proceedings of the 17th International Conference on Parallel Architectures and Compilation Techniques, 2008, pp.

\bibitem{b12} Nathan Binkert, Bradford Beckmann, Gabriel Black, Steven K. Reinhardt, Ali Saidi, Arkaprava Basu, Joel Hestness, Derek R. Hower, Tushar Krishna, Somayeh Sardashti, Rathijit Sen, Korey Sewell, Muhammad Shoaib, Nilay Vaish, Mark D. Hill, and David A. Wood. 2011. The Gem5 simulator. SIGARCH Comput. Archit. News 39, 2 (May 2011), 1–7. https://doi.org/10.1145/2024716.2024718.

\bibitem{b17} Guochu Xiong, Xiangzhong Luo, and Weichen Liu. 2025. Learning Cache Coherence Traffic for NoC Routing Design. In Great Lakes Symposium on VLSI 2025 (GLSVLSI '25), June 30--July 2, 2025, New Orleans, LA, USA. ACM, New York, NY, USA, 7 pages. https://doi.org/10.1145/3716368.3735166

\bibitem{b18} Dimitrios Chasapis, Marc Casas, Miquel Moretó, Raul Vidal, Eduard Ayguadé, Jesús Labarta, and Mateo Valero. 2015. PARSECSs: Evaluating the Impact of Task Parallelism in the PARSEC Benchmark Suite. ACM Trans. Archit. Code Optim. 12, 4, Article 41 (January 2016), 22 pages. https://doi.org/10.1145/2829952

\bibitem{b19} Huỳnh, An Helm, Christian Iwasaki, Shintaro Endo, Wataru Namsraijav, Byambajav Taura, Kenjiro. (2019). TP-PARSEC: A task parallel PARSEC benchmark suite. Journal of Information Processing. 27. 211-220. 10.2197/ipsjjip.27.211. 

\bibitem{b20} Jinchao Chen, Yang Wang, Ying Zhang, Yantao Lu, Qing Li, and Qiuhao Shu. 2025. Non-Preemptive Scheduling of Periodic Tasks with Data Dependencies in Heterogeneous Multiprocessor Embedded Systems. ACM Trans. Des. Autom. Electron. Syst. 30, 2, Article 28 (March 2025), 25 pages. https://doi-org.remotexs.ntu.edu.sg/10.1145/3711849.

\bibitem{b21} R. P. Dick, D. L. Rhodes and W. Wolf, "TGFF: task graphs for free," Proceedings of the Sixth International Workshop on Hardware/Software Codesign. (CODES/CASHE'98), Seattle, WA, USA, 1998, pp. 97-101, doi: 10.1109/HSC.1998.666245.

\bibitem{b22} M. Badr and N. E. Jerger, "SynFull: Synthetic traffic models capturing cache coherent behaviour," 2014 ACM/IEEE 41st International Symposium on Computer Architecture (ISCA), Minneapolis, MN, USA, 2014, pp. 109-120, doi: 10.1109/ISCA.2014.6853236.

\bibitem{b23} H. Chen, P. Chen, X. Luo, S. Huai and W. Liu, "LAMP: Load-Balanced Multipath Parallel Transmission in Point-to-Point NoCs," in IEEE Transactions on Computer-Aided Design of Integrated Circuits and Systems, vol. 41, no. 12, pp. 5232-5245, Dec. 2022, doi: 10.1109/TCAD.2022.3151021.

\end{thebibliography}
\end{document}